\documentclass[runningheads,fleqn]{svmult}
\usepackage{makeidx}   % allows index generation
\usepackage{graphicx}  % standard LaTeX graphics tool
\usepackage{subeqnar}  % subnumbers individual equations
\usepackage{multicol}  % used for the two-column index
\usepackage{physmult}  % flushleft layout of captions etc.
\makeindex             % used for the subject index
\usepackage{amsmath}
\usepackage{amssymb}
\begin{document}
\setcounter{chapter}{2}
\title{The Time--Energy Uncertainty Relation\label{chap:paul}
\index{time--energy uncertainty relation}\footnote{ Revised version
of Chapter 3 of the 2nd edition of the Monograph \emph{Time in
Quantum Mechanics}, eds. G. Muga \emph{et al}, Springer-Verlag,
forthcoming 2007.}}
 \toctitle{The
Time--Energy Uncertainty Relation}
\titlerunning{The Time--Energy Uncertainty Relation}

\author{Paul Busch}
\authorrunning{Paul Busch}

\institute{Department of Mathematics, University of York, York, UK}

 \maketitle

\section{Introduction}

\index{time--energy uncertainty relation}
The time--energy uncertainty relation
\begin{equation}
\Delta T\,\Delta E\geq \frac{1}{2}\hbar  \label{1paul}
\end{equation}
has been a controversial issue since the advent of quantum theory, with
respect to appropriate formalisation, validity and possible meanings.
Already the first formulations due to Bohr, Heisenberg, Pauli and
Schr\"{o}dinger are very different, as are the interpretations of the terms
used. A comprehensive account of the development of this subject up to the
1980s is provided by a combination of the reviews of Jammer \cite{Jam},
Bauer and Mello \cite{BauMel}, and Busch \cite{Bu90a,Bu90b}. More recent
reviews are concerned with different specific aspects of the subject: \cite
{Hil96,Hil98,PF}.\ The purpose of this chapter is to show that different
types of time energy uncertainty relation can indeed be deduced in specific
contexts, but that there is no unique universal relation that could stand on
equal footing with the position--momentum uncertainty relation. To this end,
we will survey the various formulations of a time energy uncertainty
relation, with a brief assessment of their validity, and along the way we
will indicate some new developments that emerged since the 1990s (Sects.
\ref{ext},\ref{int}, and \ref{obs}). In view of the existing reviews,
references to older work will be restricted to a few key sources. A
distinction of three aspects of time in quantum theory introduced in \cite
{Bu90a} will serve as a guide for a systematic classification of the
different approaches (Sect. \ref{role}).

\section{The Threefold Role of Time in Quantum Theory\label{role}}

\index{time--energy uncertainty relation}
The conundrum of the time energy uncertainty relation is related to an
ambiguity concerning the role of time in quantum theory. In the first place,
time is identified as the parameter
entering the Schr\"{o}dinger equation
and measured by an external, detached laboratory clock. This aspect will be
referred to as \emph{pragmatic}, or \emph{laboratory}, or
\emph{external time}.\index{external time}
By contrast, time as \emph{dynamical}, or \emph{intrinsic time} is
\index{intrinsic time}
defined through the dynamical behaviour of the quantum objects themselves.
Finally, time can also be considered as an observable -- called here \emph{%
observable time}, or \emph{event time}.\index{event time}
These three aspects of time in
quantum theory will be explained in some more detail.

\subsection{External Time \index{external time}}
\index{time--energy uncertainty relation}

The description of every experiment is based on a spatio--temporal
coordinatisation of the relevant pieces of equipment. For example, one will
specify the relative distances and orientations of particle sources and
detectors, as well as control the times at which external fields are
switched on and off, or record the times at which a detector fires. Such
\emph{external time} measurements are carried out with clocks that are not
dynamically connected with the objects studied in the experiment. The
resulting data are used to specify parameters in the theoretical model
describing the physical system, such as the instant or duration of its
preparation, or the time period between the preparation and the instant at
which a measurement of, say, position is performed, or the duration of a
certain measurement coupling applied.

External time is sharply defined at all scales relevant to a given
experiment. Hence there is no scope for an uncertainty interpretation with
respect to external time. However, it has been argued that the duration of
an energy measurement limits the accuracy of its outcomes. According to an
alternative proposal, the energy of an object is uncertain, or
indeterminate, during a period of preparation or measurement, since this
involves interactions. These two types of conjectured relations will be
scrutinised in Sects. \ref{aha-boh} and \ref{pre}.

\subsection{Intrinsic Time\index{intrinsic time}}
\index{time--energy uncertainty relation}

As a physical magnitude, time is defined and measured in terms of physical
systems undergoing changes, such as the straight line motion of a free
particle, the periodic circular motion of a clock dial, or the oscillations
of atoms in an atomic clock. In accordance with this observation, it can be
said that every \emph{dynamical variable} of a physical system marks the
passage of time, as well as giving an (at least approximate) quantitative
measure of the length of the time interval between two events. Hence every
nonstationary observable $A$ of a quantum system constitutes its own
characteristic time $\tau _{\varphi }\left( A\right) $ within which its mean
value changes significantly ($\varphi $ being any initial state). For
example, if $A=Q$, the position of a particle, then $\tau _{\varphi }\left(
Q\right) $ could be defined as the time it takes for the bulk of the wave
packet associated with a state vector $\varphi $ to shift by a distance
equal to the width of the packet. Or for a projection $P$, $\tau _{\varphi
}\left( P\right) $ could be the length of the greatest time interval for
which the probability $\langle \varphi _{t}|P\varphi _{t}\rangle \geq
1-\varepsilon $. Here $\varphi _{t}=e^{-itH/\hbar }\varphi $ is the state at
time $t$ in the Schr\"{o}dinger picture. Further concrete examples of
characteristic times are the time delay \index{time delay}
in scattering theory, the dwell time \index{dwell time}
in tunnelling, or the lifetime \index{lifetime}
of an unstable state
(cf. Chap. \ref{chap:gonzalo}).

The consideration of time as an entity intrinsic to the dynamical behaviour
of a physical system entails a variety of time energy uncertainty relations
in which $\Delta T$ is given by a characteristic time $\tau_{\varphi}\left(
A\right) $ associated with some dynamical variable $A$. On the other hand,
the study of dynamics often involves experimental questions about the time
of an event, the time difference between events, or the duration of a
process associated with the object system. This raises the quest for a
treatment of time as an observable.

\subsection{Observable Time\index{observable time}}
\index{time--energy uncertainty relation}

A standard experimental question in the study of decaying systems is that
about the temporal distribution of the decay events over an ensemble. More
precisely, rather than the instant of decay one will be measuring the \emph{%
time of arrival} of the decay products in a detector.
A related question is
that about the \emph{time of flight} of a particle.\index{time of flight}
Attempts to represent
these \emph{time observables} in terms of appropriate operators have been
hampered by Pauli's theorem \index{Pauli's ``theorem''}\cite{Pauli}
(cf. Chap. \ref{chap:intro}),
according to
which the semi-boundedness of any Hamiltonian $H$ precludes the existence of
a self-adjoint operator $T$ acting as a generator of a unitary group
representation of translations in the energy spectrum. In fact, the
covariance relation
\begin{equation}
e^{ihT/\hbar }He^{-ihT/\hbar }=H+hI\;,  \label{H-cov}
\end{equation}
valid for all $h\in \mathbb{R}$, immediately entails that the spectrum of $H$
should be $\mathbb{R}$. If the covariance were satisfied, it would entail
the Heisenberg canonical commutation relation, valid in a dense domain,
\begin{equation}
\left[ H,T\right] =iI\;,  \label{ccr}
\end{equation}
so that a shift generator $T$ would be canonically conjugate to the energy,
with ensuing observable--time energy uncertainty relation for any state $%
\rho $,
\begin{equation}
\Delta _{\rho }T\,\Delta _{\rho }H\geq \frac{\hbar }{2}\;.  \label{obs-ur0}
\end{equation}
In his classic paper on the uncertainty relation, Heisenberg \cite{Heis}
posited a time operator $T$ conjugate to the Hamiltonian $H$ and gave the
canonical commutation relation and uncertainty relation, without any comment
on the formal or conceptual problematics.\index{time operator}

It should be noted, however, that the Heisenberg relation is weaker than the
covariance relation; hence it is possible that the former can be satisfied
even when the latter cannot. We shall refer to operators conjugate to a
given Hamiltonian as
\emph{canonical time operators}.\index{canonical time operator}
For example, for the
harmonic oscillator Hamiltonian there do exist self-adjoint canonical time
operators $T$. In other cases, such as the free particle, symmetric
operators have been constructed which are conjugate to the Hamiltonian, but
which are not self-adjoint and do not admit self-adjoint extensions.

No general method seems to exist by which one could decide which
Hamiltonians do admit canonical, self-adjoint \emph{time }operators.
Moreover, even in cases where such time operators do not exist, there may
still be relevant experimental questions about the time of the occurrence of
an event. It is therefore appropriate to consider the approach to defining
observables in terms of the totality of statistics, that is, in terms of
positive operator valued measures (in short: POVMs).
\index{Positive Operator Valued Measure (POVM)} All standard
observables represented as self-adjoint operators are subsumed under this
general concept as special cases by virtue of their associated projection
valued spectral measures. The theory of POVMs as representatives of quantum
observables and the ensuing measurement theory are developed in \cite{BGL95}%
, including a comprehensive review of relevant literature. In Sect. \ref
{obs} we will consider examples of POVMs describing time observables and
elucidate the scope of an uncertainty relation for observable time and
Hamiltonian.\index{time--energy uncertainty relation}

In that section we will also address the important question of
interpretation of time uncertainties. The uncertainty of the decay time has
always been quoted as the prime example of the fundamental indeterminacy of
the time of occurrence of a quantum event. Yet the question remains as to
whether such an indeterminacy interpretation is inevitable, or whether the
time uncertainty is just a matter of subjective ignorance.

\section{Relation between External Time and Energy Spread\label{ext}}
\index{time--energy uncertainty relation}

One of the earliest proposed versions of a time energy uncertainty relation $%
\Delta T\,\Delta E\gtrsim h$ identifies the quantity $\Delta T$ not as an
\emph{uncertainty }but as the \emph{duration }of a measurement of energy.
The quantity $\Delta E$ has been interpreted in two ways: either as the
range within which an uncontrollable change of the energy of the object must
occur due to the measurement (starting with a state in which the energy was
more or less well defined); or as the resolution of a measurement of energy.
On the latter interpretation, if the energy measurement is repeatable, the
energy measurement resolution $\Delta E$ is also reflected in the
uncertainty of the energy in the outgoing state $\varphi $ of the object
system, that is, it is approximately equal to the root of the variance of
the Hamiltonian, $\Delta H=\left( \langle \varphi |H^{2}\varphi \rangle
-\langle \varphi |H\varphi \rangle ^{2}\right) ^{1/2}$.

The original arguments were rather informal, and this has given rise to long
controversies, leading eventually to precise quantum mechanical models on
which a decision could be based. Prominent players in this debate were Bohr,
Heisenberg and Pauli versus Einstein, with their qualitative discussions of
\emph{Gedanken }experiments; Landau and Peierls, Fock and Krylov, Aharonov
and Bohm, Kraus, Vorontsov, and Stenholm (for a detailed account, cf. \cite
{Bu90b}).

The conclusion maintained here is that an uncertainty relation between
external time duration and energy spread is not universally valid. It may
hold for certain types of Hamiltonians, but it turns out wrong in some
cases. A counter example was first provided by an energy measurement model
due to Aharonov and Bohm \cite{AhaBoh}. The debate about the validity of
this argument suffered from a lack of precise definitions of measurement
resolution and reproducibility of outcomes. This difficulty can be overcome
by recasting the model in the language of modern measurement theory using
positive operator valued measures. This analysis \cite{Bu90b} will be
reviewed and elaborated next.

\subsection{Aharonov--Bohm Energy Measurement Model\label{aha-boh}}
\index{time--energy uncertainty relation}
\index{Aharonov--Bohm energy measurement model}

We consider a system of two particles in one dimension, one particle being
the object, the other serving as a probe for a measurement of momentum. The
total Hamiltonian is given by
\[
H=\frac{P_{x}^{2}}{2m}+\frac{P_{y}^{2}}{2M}+YP_{x}g\left( t\right)\;,
\]
where $\left( X,P_{x}\right) $ and $\left( Y,P_{y}\right) $ are the position
and momentum observables of the object and probe, respectively, and $m,M$
are their masses. The interaction term produces a coupling between the
object momentum $P_{x}$ to be measured and the momentum $P_{y}$ of the probe
as the read-out observable. The function $g\left( t\right) $ serves to
specify the duration and strength of the interaction as follows:
\[
g\left( t\right) =\left\{
\begin{array}{ll}
g_{0}\;\;\; & {\rm{if\ \ }}0\leq t\leq \Delta t\;,\\
0 & {\rm{otherwise\;.}}
\end{array}
\right.
\]
The Heisenberg equations for the positions and momenta read
\[
\begin{array}{lll}
\dot{X}=\frac{1}{m}P_{x}+Yg\left( t\right) \,, &  & \dot{P}_{x}=0, \\
&  &  \\
\dot{Y}=\frac{1}{M}P_{y}\,, &  & \dot{P}_{y}=-P_{x}g\left( t\right) \;.
\end{array}
\]
This is solved as follows:
\[
P_{x}=P_{x}^{0}\,,\;\;\;P_{y}=P_{y}^{0}-P_{x}^{0}\,g_{0}\,\Delta t\,,\;\;\;\;%
{\rm{for \ }}t\geq \Delta t\;.
\]
The kinetic energy of the object before and after the interaction is given
by one and the same operator:
\[
H_{0}=\frac{m}{2}\dot{X}^{2}=\frac{P_{x}^{2}}{2m}\;.
\]
Thus, the value of kinetic energy $H_{0}$ can be obtained by determining the
momentum $P_{x}$ in this measurement. During the interaction period the
kinetic energy $\frac{m}{2}\dot{X}^{2}$ varies but the first moments before
and after the measurement are the same. This is an indication of a \emph{%
reproducible} energy measurement. Following Aharonov and Bohm, one could
argue that achieving a given resolution $\Delta p_{x}$ requires the change
of deflection of the probe $\Delta \left( P_{y}-P_{y}^{0}\right) $ due to a
shift of the value of $P_{x}$ of magnitude $\Delta p_{x}$ to be greater than
the initial uncertainty of the probe momentum, $\Delta P_{y}^{0}$. This
yields the following threshold condition: \index{time--energy uncertainty
relation}
\[
\Delta p_{x}\,g_{0}\,\Delta t\cong \Delta P_{y}^{0}\;.
\]
By making $g_{0}$ large enough, ``both $\Delta t$ and $\Delta p_{x}$ can be
made arbitrarily small for a given $\Delta P_{y}^{0}\,$'' \cite{AhaBoh}.

This is the core of Aharonov and Bohm's refutation of the external time
energy uncertainty relation: the energy measurement can be made in an
arbitrarily short time and yet be reproducible and arbitrarily accurate.

It is instructive to reformulate the whole argument within the
Schr\"{o}dinger picture, as this will allow us to find the POVMs
\index{Positive Operator Valued Measure (POVM)} for
momentum and kinetic energy associated with the relevant measurement
statistics. The property of reproducibility presupposes a notion of
initially relatively sharp values of the measured observable. We take the
defining condition for this to be the following: the uncertainty of the
final probe momentum is approximately equal to the initial uncertainty. Let $%
\Phi=\varphi\otimes\phi$ be the total Heisenberg state of the object ($%
\varphi$) plus probe ($\phi$). The final probe momentum variance is found to
be
\[
\left( \Delta_{\Phi}P_{y}\right) ^{2}=\left( \Delta_{\phi}P_{y}^{0}\right)
^{2}+g_{0}^{2}\,\Delta t^{2}\,\left( \Delta_{\varphi}P_{x}\right) ^{2}\;.
\]
Sharpness of the object momentum corresponds to the last term being
negligible.

First we calculate the probability of obtaining a value $P_{y}$
in an interval $S$. The corresponding spectral projection will be denoted $%
E^{P_{y}}\left( S\right) $.
The following condition determines the POVM
\index{Positive Operator Valued Measure (POVM)} of
the measured \emph{unsharp }momentum observable of the object:
\[
\langle \Phi _{\Delta t}|I\otimes E^{P_{y}}\left( S\right) \Phi _{\Delta
t}\rangle =\langle \varphi |A\left( S\right) \varphi \rangle \;\;\;{\rm{for
\, all \ }}\varphi\;,
\]
where $\Phi _{\Delta t}=\exp \left( -i\Delta t\,H/\hbar \right) \,\varphi
\otimes \phi $ is the total state immediately after the interaction period,
i.e., \index{time--energy uncertainty relation}
\[
\Phi _{\Delta t}\left( p_{x},p_{y}\right) =e^{ -i p_x^2\,\Delta t/2m\hbar
-i\gamma \left( p_{x},p_{y},\Delta t\right) /\hbar } \,\varphi \left( p_{x}\right) \,\phi \left(
p_{y}+p_{x}g_{0}\Delta t\right)\;,
\]
\[
\gamma \left( p_{x},p_{y},\Delta t\right) =\frac{1}{6M}\,p_{x}^{2}\,g_{0}%
\,\Delta t^{3}+\frac{1}{2M}\,p_{x}\,p_{y}\,g_{0}\,\Delta t^{2}+\frac{1}{2M}%
\,p_{y}^{2}\,\Delta t\;.
\]
One obtains:
\[
A\left( S\right) =E_{f}^{P_{x}}\left( -\frac{S}{g_{0}\Delta t}\right)\;,
\]
which is an \emph{unsharp momentum} observable ($\ast $ denoting
convolution),
\begin{eqnarray}
E_{f}^{P_{x}}\left( R\right) & =&\chi _{R}\ast f\left( P_{x}\right) =\int_{%
\mathbb{R}}dp\,f\left( p\right) \,E^{P_{x}}\left( R+p\right)\;,  \label{p-pov}
\\
f\left( p\right) & =&g_{0}\,\Delta t\,\left| \phi \left( p\,g_{0}\,\Delta
t\right) \right| ^{2}\;.
\label{p-confid}
\end{eqnarray}
Due to the properties of the convolution it is straightforward to verify
that these positive operators form a POVM,
\index{Positive Operator Valued Measure (POVM)} that is,
(countable) additivity
over disjoint sets and normalisation $E_{f}^{P_{x}}\left( \mathbb{R}\right)
=I$ are satisfied. It is thus seen that the resolution of the measurement,
described by the confidence distribution $f$, is determined by the initial
probe state as well as the interaction parameter $g_{0}\Delta t$. In fact, a
measure of the inaccuracy is given by the width of the distribution $f$,
which can be characterised (for suitable probe states $\phi $) by the
variance:
\begin{equation}
(\Delta p_{x})^2=\mathrm{Var}_{f}\left( p\right) =\left( \frac{1}{g_{0}\Delta
t}%
\right) ^{2}\,\mathrm{Var}_{\phi }\left( P_{y}\right)\;.  \label{p-inacc}
\end{equation}
It is clear that increasing the parameter $g_{0}\Delta t$ leads to a more
and more sharply peaked function $f$. This is to say that the inaccuracy of
the momentum measurement, given by the width $\Delta f$ of $f$, can be
arbitrarily increased for any fixed value of the duration $\Delta t$. The
same will be seen to be true for the inaccuracy of the measured values of
energy inferred from this momentum measurement. This disproves the \emph{%
inaccuracy }version of the external--time energy uncertainty relation where $%
\Delta E$ is taken to be the energy measurement inaccuracy.

In order to assess the reproducibility properties of the measurement, we
need to investigate the state change of the object due to the measurement.
The final object state $\rho _{R}$ conditional upon an outcome $p_{x}$ in $R$
is determined via the following relation: for all states $\varphi $ and all
object operators $a$, \index{time--energy uncertainty relation}
\[
{\rm{tr}}\left[ a\,\rho _{R}\right] =\langle \Phi _{\Delta t}|a\otimes
E^{P_{y}}\left( -Rg_{0}\Delta t\right) \Phi _{\Delta t}\rangle\;.
\]
One obtains:
\[
\rho _{R}=\int_{R}dp_{x}^{\prime }\,A_{p_{x}^{\prime }}\,|\varphi \rangle
\langle \varphi |\,A_{p_{x}^{\prime }}^{\ast }
\]
where the operators $A_{p_{x}^{\prime }}$ act as
\begin{eqnarray*}
\left( A_{p_{x}^{\prime }}\varphi \right) \left( p_{x}\right) & =&\left(
g_{0}\Delta t\right) ^{1/2}\,e^{  -ip_{x}^{2}\Delta t/2m\hbar }
\,e^{ -i\gamma ( p_{x},-p_{x}^{\prime }g_{0}\Delta t,\Delta
t) /\hbar } \,\times \\
&& \qquad\qquad\qquad\qquad\qquad\qquad%\;\;\;\;\;\;\;\;\;\;\;\;\;\;\ \ \ \;\;\;\;\;\;\;\;\;\;\;\;\;\;\;
\times \phi
\left( \left( p_{x}-p_{x}^{\prime }\right) g_{0}\Delta t\right) \,\varphi
\left( p_{x}\right)\;.
\end{eqnarray*}
\index{time--energy uncertainty relation}
The momentum distribution is (up to normalisation):
\[
\langle p_{x}|\rho _{R}|p_{x}\rangle =\int_{R}dp_{x}^{\prime }\left| \left(
A_{p_{x}^{\prime }}\varphi \right) \left( p_{x}\right) \right| ^{2}=\chi
_{X}\ast f\left( p_{x}\right) \,\left| \varphi \left( p_{x}\right) \right|
^{2}\;.
\]
If $\left| \varphi \left( p_{x}\right) \right| ^{2}$ is sharply peaked at $%
p_{x}^{0}$, in the sense that
\[
\left| \phi \left( \left( p_{x}-p_{x}^{\prime }\right) g_{0}\Delta t\right)
\right| ^{2}\left| \varphi \left( p_{x}\right) \right| ^{2}\cong \left| \phi
\left( \left( p_{x}^{0}-p_{x}^{\prime }\right) g_{0}\Delta t\right) \right|
^{2}\left| \varphi \left( p_{x}\right) \right| ^{2}\;,
\]
then one has
\begin{equation}
\langle p_{x}|\rho _{R}|p_{x}\rangle \cong \chi _{R}\ast f\left(
p_{x}^{0}\right) \,\left| \varphi \left( p_{x}\right) \right| ^{2}\;.
\label{p-reprod}
\end{equation}
Hence if $\varphi $ is such a \emph{near-eigenstate} of $P_{x}$, then the
conditional final state has practically the same sharply peaked momentum
distribution. In other words, the present model practically preserves
near-eigenstates. It follows indeed that the measurement allows one to
determine the kinetic energy with negligible disturbance of any pre-existing
(approximately sharp) value. Thus the \emph{disturbance }version of the
purported external--time energy uncertainty relation is ruled out.

We show next in which sense the above momentum measurement scheme serves as
a measurement of kinetic energy. In fact the relation $H_{0}=P_{x}^{2}/2m$
translates into the following functional relationship between the spectral
measures of $H_{0}$ and $P_{x}$: we have
\[
H_{0}=\frac{P_{x}^{2}}{2m}=\int_{-\infty }^{+\infty }\frac{p^{2}}{2m}%
E^{P_{x}}\left( dp\right) =\int_{0}^{+\infty }e\,E^{H_{0}}\left( de\right)\;,
\]
and so
\[
E^{H_{0}}\left( Z\right) =E^{P_{x}}\left( h^{-1}\left( Z\right) \right)
,\;\;\;Z\subseteq \mathbb{R}^{+},\;\;\;h\left( p\right) =\frac{p^{2}}{2m}\;.
\]
This suggests that in the above unsharp momentum measurement, one should
record such subsets $R$ of the momentum spectrum which are images of some $%
Z\subseteq \mathbb{R}^{+}$ under the map $h^{-1}$. This leads to the
following positive operators which constitute a POVM
\index{Positive Operator Valued Measure (POVM)}
on $\mathbb{R}^{+}$:
\[
E_{f}^{H_{0}}\left( Z\right) :=E_{f}^{P_{x}}\left( h^{-1}\left( Z\right)
\right) =\int_{\mathbb{R}}f\left( p\right) \,E^{P_{x}}\left( h^{-1}\left(
Z\right) +p\right) \,dp\;.
\]
Let us assume the confidence function $f$ is inversion symmetric, $f\left(
-p\right) =f\left( p\right) $. Then, since the set $h^{-1}\left( Z\right) $
is inversion symmetric, the convolution $\chi _{h^{-1}\left( Z\right) }\ast f
$ also shares this property. Hence the positive operators $%
E_{f}^{H_{0}}\left( Z\right) $ are actually functions of $H_{0}$ and
constitute a smearing of the spectral measure of $H_{0}$: \index{time--energy
uncertainty relation}
\begin{eqnarray}
E_{f}^{H_{0}}\left( Z\right) & =&\chi _{h^{-1}\left( Z\right) }\ast f\left(
P_{x}\right) =\chi _{h^{-1}\left( Z\right) }\ast f\left( \left(
2mH_{0}\right) ^{1/2}\right)   \label{H-pov}
\\
& =&\int_{Z}\left( \frac{m}{2e}\right) ^{1/2}\,f\left( \left( 2mH_{0}\right)
^{1/2}-\left( 2me\right) ^{1/2}\right) \,de\;.
\nonumber
%\notag
\end{eqnarray}
This is a corroboration of the fact that the unsharp momentum measurement
constitutes an unsharp measurement of energy. The expected readings and
their variances are obtained as follows:
\[
\langle p^{n}\rangle _{f}=\int_{\mathbb{R}}p^{n}f\left( p\right) dp=\left(
\frac{1}{g_{0}\Delta t}\right) ^{n}\,\langle P_{y}^{n}\rangle _{\phi }\;,
\]
then
\begin{equation}
\langle H_{0}\rangle _{\varphi ,f}=\langle \varphi |\int_{0}^{\infty
}e\,E_{f}^{H_{0}}\left( de\right) \varphi \rangle =\langle H_{0}\rangle
_{\varphi }+\left( \frac{1}{g_{0}\Delta t}\right) ^{2}\mathrm{\,}%
\left\langle \frac{P_{y}^{2}}{2m}\right\rangle _{\phi }\;,  \label{H-val}
\end{equation}
and
\begin{eqnarray}
&&\mathrm{Var}_{\varphi ,f}\left( H_{0}\right) =\langle
H_{0}^{2}\rangle _{\varphi ,f}-\left( \langle H_{0}\rangle _{\varphi
,f}\right) ^{2} \label{H-var}
\\
&&= \mathrm{Var}_{\varphi }\left( H_{0}\right) +\left( \frac{1}{g_{0}\Delta t}%
\right) ^{4}\,\mathrm{Var}_{\phi }\left( \frac{P_{y}^{2}}{2m}\right)
+4\left( \frac{1}{g_{0}\Delta t}\right) ^{2}\,\langle H_{0}\rangle _{\varphi
}\,\left\langle \frac{P_{y}^{2}}{2m}\right\rangle _{\phi }\;.
\nonumber
%\notag
\end{eqnarray}
There is a distortion of the expectation values towards slightly larger
values, and the energy measurement inaccuracy is measured by the last two
terms in the last equation. Both the distortion as well as the accuracy can
be made arbitrarily small by choosing a suitably large coupling parameter $%
g_{0}$, although it must be noted that the inaccuracy depends on the value
of the object energy.

We conclude therefore, in agreement with Aharonov and Bohm, that a
reproducible energy measurement is possible with arbitrary accuracy and
within arbitrarily short time.
\index{time--energy uncertainty relation}

However, very recently Aharonov and Reznik \cite{AhaRez} have taken up the
issue again, considering this time energy measurements carried out from
\emph{within} the system. In this situation the conclusion is that due to a
back-reaction of the energy measurement on the internal clock, an accuracy $%
\delta E$ requires the duration $\tau _{0}$, measured internally, to be
limited by the uncertainty relation
\begin{equation}
\tau _{0}\,\delta E\geq \hbar \;.  \label{AR-ur}
\end{equation}
What is actually shown in the analysis of \cite{AhaRez} is that the clock
rate is uncertain and hence the duration has an \emph{uncertainty} $\Delta
\tau _{0}\geq \hbar /\delta E$. This conclusion is in accordance with the
quantum clock uncertainty relation which will be presented in Sect. \ref
{q-clock}.

\subsection{Relation between Preparation Time and Energy\label{pre}}
\index{preparation time and energy}
\index{time--energy uncertainty relation}

An uncertainty relation for the indeterminacy of the energy of a system and
the duration of an external perturbation has been proposed and accepted as
valid even by opponents to the external--time energy relation (cf. the
review of Bauer and Mello \cite{BauMel}). The duration of the perturbation
is defined \emph{dynamically} as the approximate time period during which
the interaction energy is non-negligible. Hence this type of time energy
uncertainty relation is best classified as one associated with dynamical
time, although in a measurement context the duration of interaction is fixed
with reference to a laboratory clock. A particular instance of this type of
uncertainty relation occurs in the preparation of a quantum system: the
interaction with the preparation devices can be regarded as an external
perturbation, so that one may note:
\begin{equation}
T_{\mathrm{prep}}\,\Delta E\gtrsim \hbar \;,  \label{prep-ur}
\end{equation}
where $T_{\mathrm{prep}}$ denotes the duration of the preparation
(perturbation) and $\Delta E$ is some suitable measure of the width of the
energy distribution, such as those introduced in the next section.

This preparation time relation has been deduced by Moshinsky \cite{Mosh} in
an exactly soluble potential model of the preparation of a particle by means
of a slit with a shutter which is opened during a time interval $T_{\mathrm{%
prep}}$. \index{Moshinsky's shutter}
This time period determines the width of the Bohr--Wigner time of
passage distribution (cf. Subsection \ref{BWur} below, equation (\ref{Wig-ft}%
)), whereas the energy uncertainty $\Delta E$ is given by the width of the
energy distribution of the outgoing particle, given a sharp initial energy $%
E_{0}$:
\[
\mathfrak{p}\left( E:E_{0},T_{\mathrm{prep}}\right) \propto E^{1/2}\,\frac
{\sin^{2}\left( \left( E-E_{0}\right) T/2\hbar\right) }{\left(
E-E_{0}\right) ^{2}}\;.
\]
Similar distributions are known to arise for the short-time energy
distribution of a decaying state as well as in first-order perturbation
theory. We conclude that it is impossible to simultaneously prepare a sharp
energy and a sharp time of passage. This is an indication of the
complementarity of event time and energy.

A relation of the form (\ref{prep-ur}) was derived in a somewhat
different context by Partovi and Blankenbecler \cite{PaBl1986}; they
showed that the most likely state compatible with the probability
distributions of the position of a free particle measured at two
times with separation $T$ has an energy dispersion that must satisfy
(\ref{prep-ur}). These authors interpret the time interval $T$
between the two measurement as the duration of a multi-time
measurement whose aim it is to estimate the state that gives rise to
the statistical data obtained.

\section{Relations Involving Intrinsic Time\label{int}}

In this section we review different ways of quantifying measures of times
that are intrinsic to the system and its evolution.

\subsection{Mandelstam--Tamm Relation}
\index{Mandelstam--Tamm relation}
\index{time--energy uncertainty relation}

A wide class of measures of intrinsic times has been provided by Mandelstam
and Tamm \cite{MT}. An elegant formulation of the ensuing universal \emph{%
dynamical}, or \emph{intrinsic--}time energy uncertainty relations was given
in the textbook of Messiah. Let $A$ be a non-stationary observable.
Combining the Heisenberg equation of motion for $A$,
\begin{equation}
i\hbar \,\frac{dA}{dt}=AH-HA\;,  \label{Heis-pic}
\end{equation}
with the general uncertainty relation
\begin{equation}
\Delta _{\rho }A\,\Delta _{\rho }H\geq \frac{1}{2}\left| \langle
AH-HA\rangle _{\rho }\right|\;,   \label{ur}
\end{equation}
and introducing the characteristic time
\begin{equation}
\tau _{\rho }\left( A\right) =\frac{\Delta _{\rho }A}{\left| \frac{d}{dt}%
\langle A\rangle _{\rho }\right| }  \label{MS-tau}
\end{equation}
(whenever the denominator is nonzero), one obtains the inequality
\begin{equation}
\tau _{\rho }\left( A\right) \,\Delta _{\rho }H\geq \frac{1}{2}\hbar\;.
\label{MT-ur}
\end{equation}
Here we have used the notation $\langle X\rangle _{\rho }=\mathrm{tr}\left[
\rho X\right] $, $\left( \Delta _{\rho }X\right) ^{2}=\langle X^{2}\rangle
_{\rho }-\langle X\rangle _{\rho }^{2}$.

As an illustration we consider the case of a free particle. Let $A=Q$ be the
particle position and let $\rho $ be a pure state represented by a unit
vector $\varphi $. Assume the momentum $P$ is fairly sharply defined in that
state, i.e., $\Delta _{\varphi }P\ll \left| \langle P\rangle _{\varphi
}\right| $. Now the time derivative of position is the velocity, $d\langle
Q\rangle _{\varphi }/dt=\langle P\rangle _{\varphi }/m=\langle V\rangle
_{\varphi }$, so we have
\begin{equation}
\tau _{\varphi }\left( Q\right) =\frac{\Delta _{\varphi }Q}{\left| \langle
V\rangle _{\varphi }\right| }\;.  \label{MT-tau-pos}
\end{equation}
From the Schr\"{o}dinger equation for a free particle we have
$$
\left( \Delta _{\varphi }Q\right) ^{2}=\left( \Delta _{\varphi }Q\left(
0\right) \right) ^{2}+\left( \Delta _{\varphi }V\right) ^{2}t^{2}+\left\{
\langle Q\left( 0\right) V+VQ\left( 0\right) \rangle _{\varphi }-2\langle
V\rangle _{\varphi }\langle Q\left( 0\right) \rangle _{\varphi }\right\}
t\,.
$$
Using the uncertainty relation in the general form\index{time--energy
uncertainty relation}
\begin{eqnarray*}
\left( \Delta _{\varphi }Q\right) ^{2}\,\left( \Delta _{\varphi
}P\right) ^{2} &\geq& \frac{1}{4}\left| \langle Q\left( 0\right)
P-PQ\left( 0\right)
\rangle _{\varphi }\right| ^{2} \\
 %\;\;\;\;\;\;\;\;\;\;\;\;\;\;\;\;\;\;\;\;\;\;\;\;\;\;\;\;\;\;\;\;\;\;
 &&\qquad+\frac{1}{4}\left\{ \langle Q\left( 0\right)
V+VQ\left( 0\right) \rangle _{\varphi }-2\langle V\rangle _{\varphi }\langle
Q\left( 0\right) \rangle _{\varphi }\right\} ^{2},
\end{eqnarray*}
we find the estimate
\[
\left( \Delta _{\varphi }Q\right) ^{2}\leq \left( \Delta _{\varphi }Q\left(
0\right) +t\,\Delta _{\varphi }V\right) ^{2}\;.
\]
Putting $t=\tau _{\varphi }\left( Q\right) $, this gives
\[
\Delta _{\varphi }Q\leq \Delta _{\varphi }Q\left( 0\right) \,\left[ 1+\frac{%
\Delta _{\varphi }P}{\left| \langle P\rangle _{\varphi }\right| }\right]
^{1/2}\cong \Delta _{\varphi }Q\left( 0\right) .
\]
This estimate follows from the assumption of small variance for $P$ and this
corresponds to the limiting case of slow wave packet spreading. Thus the
characteristic time $\tau _{\varphi }\left( Q\right) $ is indeed seen to be
the period of time it takes the wave packet to propagate by a distance equal
to its width. It can also be said that this is the approximate time for the
packet to pass a fixed point in space. Insofar as the position of the
particle is indeterminate within approximately $\Delta _{\varphi }Q\left(
0\right) $ one may be tempted to interpret this characteristic time as the
indeterminacy of the time of passage.\index{time of passage}
The event `particle passes a point $%
x_{0}$' has an appreciable probability only within a period of duration $%
\tau _{\varphi }\left( Q\right) $.

\subsection{Lifetime of a Property}
\index{lifetime}
\index{time--energy uncertainty relation}

Let $P$ be a projection, $U_{t}=\exp \left( -itH/\hbar \right) $, $\psi _{0}$
be a unit vector representing the state of a quantum system. We consider the
function
\begin{equation}
\mathfrak{p}\left( t\right) =\langle \psi _{0}|U_{t}^{-1}PU_{t}\psi _{0}\rangle .
\label{pt}
\end{equation}
The Mandelstam--Tamm relation
\index{Mandelstam--Tamm relation}
yields:
\[
\left| \frac{d\mathfrak{p}}{dt}\right| \leq \frac{2}{\hbar }\Delta _{\psi
_{0}}H\,\left[ \mathfrak{p}\left( 1-\mathfrak{p}\right) \right] ^{1/2}\;.
\]
Integration of this inequality with the initial condition $\mathfrak{p}\left(
0\right) =1$ yields
\begin{equation}
\mathfrak{p}\left( t\right) \geq \cos ^{2}\left( t\,\Delta _{\psi _{0}}H/\hbar
\right) ,\;\;\;0\leq t\leq \frac{\pi }{2}\,\frac{\hbar }{\Delta _{\psi _{0}}H%
}\equiv t_{0}\;.  \label{MT-p}
\end{equation}
The initial condition means that the property $P$ was actual in the state $%
\psi _{0}$ at time $t=0$. One may define the lifetime $\tau _{P}$ of the
property $P$ by means of the condition $\mathfrak{p}\left( \tau _{P}\right) =%
\frac{1}{2}$. Hence one obtains the uncertainty relation
\begin{equation}
\tau _{P}\,\Delta _{\psi _{0}}H\geq \frac{\pi \,\hbar }{4}\;.
\label{MT-lifetime}
\end{equation}
This relation was derived by Mandelstam and Tamm for the special case of $%
P=|\psi _{0}\rangle \langle \psi _{0}|$.

There are alternative approaches to defining the lifetime of a state and
obtaining an energy-time uncertainty relation for the lifetime. For
example, Grabowski \cite{Grabo} defines
\begin{equation}
\tau _{0}=\int_{0}^{\infty }\mathfrak{p}\left( t\right) \,dt\;,  \label{Grabo-tau}
\end{equation}
which yields
\begin{equation}
\tau _{0}\,\Delta _{\psi _{0}}H\geq \frac{\hbar }{2}  \label{Grabo-lifetime}
\end{equation}
provided the Hamiltonian has no singular continuous spectrum.

The variance of $H$ may be infinite in many situations, so that the above
relations are of limited use. We will review below a variety of approaches
based on alternative measures of the width of the energy distribution in a
state $\psi_{0}$.

\subsection{Bohr--Wigner Uncertainty Relation\label{BWur}}
\index{Bohr--Wigner uncertainty relation}
\index{time--energy uncertainty relation}

Fourier analysis gives `uncertainty' relations for any wave propagation
phenomenon in that it gives a reciprocal relationship between the widths of
the spatial/temporal wave pattern on one hand, and the wave
number/frequency distributions on the other. On the basis of this classical
wave analogy, Bohr \cite{Bohr28p} proposed a time energy uncertainty relation
which appeared to assume the same status as the corresponding
position/momentum relation:
\begin{equation}
\Delta t\,\Delta E\gtrsim h,\;\;\;\Delta x\,\Delta p\gtrsim h\;.
\label{Bohr-ur}
\end{equation}
Hilgevoord \cite{Hil96} presents a careful discussion of the sense in which
a treatment of time and energy variables on equal footing to position and
momentum variables is justified.

A more formal approach in this spirit was pursued by Wigner \cite{Wig72},
who considered a positive temporal distribution function associated with the
wave function $\psi$ of a particle:
\begin{equation}
\mathfrak{p}_{x_{0}}\left( t\right) =|f\left( t\right) |^{2},\;\;\;\;\;f\left(
t\right) =\psi\left( x_{0},t\right)\;.  \label{Wig-pxt}
\end{equation}
In the limit $\Delta_{\psi}P\ll\left| \langle P\rangle_{\psi}\right| $, the
width of this distribution is of the order of $\tau_{\psi}\left( Q\right) $.
The quantity $\Delta E$ measures the width of the Fourier transform $\tilde
{f}$ of $f$.

This method can be extended to other types of characteristic times. Define
\begin{equation}
f\left( t\right) =\langle \varphi |\psi _{t}\rangle ,\;\;\;\tilde{f}\left(
E\right) =\left( 2\pi \right) ^{-1}\int_{-\infty }^{\infty }f\left( t\right)
\,e^{itE/\hbar }\,dt\;,  \label{Wig-ft}
\end{equation}
and the moments (providing the denominators are finite)
\begin{equation}
\langle t^{n}\rangle _{f}=\frac{\int_{0}^{\infty }\left| f\left( t\right)
\right| ^{2}\,t^{n}\,dt}{\int_{0}^{\infty }\left| f\left( t\right) \right|
^{2}\,\,dt}\;,\;\;\;\langle E^{n}\rangle _{\tilde{f}}=\frac{\int_{0}^{\infty
}\left| \tilde{f}\left( E\right) \right| ^{2}\,E^{n}\,dE}{\int_{0}^{\infty
}\left| \tilde{f}\left( E\right) \right| ^{2}\,\,dE}\;.  \label{Wig-moments}
\end{equation}
The previous case considered by Bohr is formally included by replacing $%
| \varphi \rangle $ with $|x_{0}\rangle $, an improper position eigenstate.
One obtains an uncertainty relation for the variances $\left( \Delta
_{f}t\right) ^{2}=\langle t^{2}\rangle _{f}-\langle t\rangle _{f}^{2}$, $%
\left( \Delta _{\tilde{f}}E\right) ^{2}=\langle E^{2}\rangle _{\tilde{f}%
}-\langle E\rangle _{\tilde{f}}^{2}$:
\begin{equation}
\Delta _{f}t\,\Delta _{\tilde{f}}E\geq \frac{\hbar }{2}\;.\label{Wig-ur}
\end{equation}
It must be noted that neither of the distributions $\left| f\left( t\right)
\right| ^{2}$ and $\left| \tilde{f}\left( E\right) \right| ^{2}$ is
normalised, nor will they always be normalisable. Moreover, their
operational meaning is not immediately obvious. The following is a possible,
albeit indirect, way of associating these distributions with physical
measurements.\index{time--energy uncertainty relation}

Assume the state $\psi $ is prepared at time $t=0$, and that at time $t>0$ a
repeatable measurement of energy is made and found to give a value in a
small interval $Z$ of width $\delta E$ and centre $E_{0}$, after which a
measurement of the property $P_{\varphi }=|\varphi \rangle \langle \varphi |$
is made. We calculate the probability for this sequence of events, under the
assumption that $H$ has a nondegenerate spectrum with improper eigenstates $%
| E\rangle $:
\begin{eqnarray}
\mathfrak{p}& =&\mathfrak{p}_{\psi }\left( E^{H}\left( Z\right) ,P_{\varphi }\right) =%
\mathrm{tr}\left[ P_{\varphi }\,E^{H}\left( Z\right) \,e^{-itH/\hbar
}\,|\psi \rangle \langle \psi |\,e^{itH/\hbar }\,E^{H}\left( Z\right)
\,P_{\varphi }\right]
\nonumber
\\
& =&\int_{Z}dE\int_{Z}dE^{\prime }\,\left\langle \varphi |E\right\rangle
\,\left\langle E^{\prime }|\varphi \right\rangle \,\left\langle \psi
| E^{\prime }\right\rangle \,\left\langle E|\psi \right\rangle \,e^{-it\left(
E-E^{\prime }\right) /\hbar }.  \label{Wig-pt}
\end{eqnarray}
Assuming that $Z$ is sufficiently small so that the functions $\left\langle
E|\psi \right\rangle $ and $\left\langle E|\varphi \right\rangle $ are
practically constant within $Z$, we have:
\begin{equation}
\mathfrak{p}\cong \left| \left\langle E_{0}|\varphi \right\rangle \right|
^{2}\,\left| \left\langle E_{0}|\psi \right\rangle \right| ^{2}\,\left(
\delta E\right) ^{2}\cong \left| \tilde{f}\left( E_{0}\right) \right|
^{2}\,\left( \delta E\right) ^{2}\;.  \label{Wig-pt2}
\end{equation}

As an illustration we reproduce the standard formulas for the exponential
decay law.\index{exponential decay}
This is known to hold in an intermediate time range, while
deviations must occur for short as well as long times, see Sects. \ref{stb}
and \ref{ltb}.
The Mandelstam--Tamm \index{Mandelstam--Tamm relation}
relation for the lifetime of a property already indicates that the
short-time behaviour of the survival probability\index{survival probability}
is a power law $1-\mathfrak{p}%
\propto t^{2}$.

For $H$ with nondegenerate spectrum, one has
\begin{eqnarray}
f\left( t\right) &=&\left\langle \psi _{0}|\psi _{t}\right\rangle
=\int_{-\infty }^{\infty }\,e^{  -itE/\hbar } \,\tilde{f}\left(
E\right) \,dE
%\notag
\nonumber
\\
&& \cong \exp \left( -\left| t\right| \left( \Gamma /2\hbar \right)
-itE_{0}/\hbar \right) ,  \label{ft-expon} \\
&&
%\notag
\nonumber
\\
\tilde{f}\left( E\right) & =&\left| \left\langle E|\psi _{0}\right\rangle
\right| ^{2}\cong \frac{1}{\pi }\,\frac{\Gamma /2}{\left( E-E_{0}\right)
^{2}+\left( \Gamma /2\right) ^{2}}\;.  \label{f-til-E-Lor}
\end{eqnarray}
The Lorentzian distribution $\tilde{f}\left( E\right) $ has no finite
variance, hence as an alternative measure of the energy spread one usually
takes the full width at half-height, $\delta E=\Gamma $. The lifetime $\tau $
\index{lifetime}
of the state $\psi _{0}$ is defined via \index{time--energy uncertainty
relation}
\begin{equation}
\mathfrak{p}\left( \tau \right) =e^{ -\tau \Gamma /\hbar } =1/e\;,
\label{p-life-expon}
\end{equation}
so that one obtains the famous lifetime-linewidth relation
\index{lifetime-linewidth relation}
\begin{equation}
\tau \,\Gamma =\hbar\;.  \label{life-line-ur}
\end{equation}
One can also use the Wigner measures, which are
\begin{equation}
\Delta _{f}t=\frac{\hbar }{2\Gamma }=\sqrt{2\tau }\,,\;\;\;\Delta _{\tilde{f}%
}E=\Gamma /2\;.  \label{Wig-lifetime}
\end{equation}
It must be noted that here the relevant distribution is $\left| \tilde{f}%
\left( E\right) \right| ^{2}=\left| \left\langle E|\psi _{0}\right\rangle
\right| ^{4}$. Hence we have:
\begin{equation}
\Delta _{f}t\,\Delta _{\tilde{f}}E=\frac{\sqrt{2}}{2}\,\hbar\;.
\label{Wig-life-ur}
\end{equation}

A novel application of a Wigner-type uncertainty relation has been proposed
recently \cite{Muga1} which identifies $\tilde{f}(E)$ as the energy
amplitude of a state, in which case the associated $\left| f\left( t\right)
\right| ^{2}$ is found to coincide with the time of arrival distribution due
to Kijowski \cite{Kijo}.\index{Kijowski's distribution}

Another approach to defining a formal probability distribution for
time based on the statistics of measurements of a time dependent
observable $A$ was attempted by Partovi and Blankenbecler
\cite{PaBl1986}. This approach presupposes that the time dependence
of the expectation $A(t):={\rm tr}[\rho(t) A]$ is strictly
monotonic. It seems that the scheme of a proof of a time-energy
uncertainty relation for the dispersion of the ensuing time
distribution provided in \cite{PaBl1986} gives tangible results
essentially when the (self-adjoint) operator $A$ satisfies the
canonical commutation relation with the Hamiltonian, which is known
to be possible only in very special cases.

\subsection{Further Relations Involving Intrinsic Time}
\index{time--energy uncertainty relation}

In more realistic models of decaying systems, the measures of spread
introduced in the previous section turn out inadequate. Bauer and Mello \cite
{BauMel} have studied alternative measures with a wider scope of
applications. For example, they define a concept of \emph{equivalent width},
given by
\begin{equation}
W\left( \phi \right) =\left( \phi \left( x_{0}\right) \right)
^{-1}\,\int_{-\infty }^{\infty }\phi \left( x\right) \,dx
\label{BM-equiv-width}
\end{equation}
whenever the right hand side is well defined. They then prove that the
following relation holds:
\begin{equation}
W\left( \phi \right) \,W\left( \tilde{\phi}\right) =2\pi \hbar\;.
\label{equiv-width-ur}
\end{equation}
In the case of a decaying state,
\[
\tilde{\phi}\left( E\right) =\left| \tilde{f}\left( E\right) \right|
^{2}=\left| \left\langle E|\psi _{0}\right\rangle \right| ^{4}\;,
\]
so that the inverse Fourier transform turns out to be the autocorrelation
function of $f$:
\[
\phi \left( t\right) =\int_{-\infty }^{\infty }\,e^{  -itE/\hbar}
 \tilde{\phi}\left( E\right) dE=\frac1{2\pi \hbar }\int_{-\infty }^{\infty }\overline{f}\left( t^{\prime }\right)
\,f\left( t+t^{\prime }\right) \,dt^{\prime }=\overline{f}\times f\left(
t\right)\,.
\]
On proving the inequality $\left| W\left( \overline{f}\times f\right)
\right| \leq W\left( \left| f\right| \times \left| f\right| \right) $, one
obtains a time energy uncertainty relation for equivalent widths:
\begin{equation}
W\left( \left| f\right| \times \left| f\right| \right) \,W\left( \left|
\tilde{f}\right| ^{2}\right) \geq 2\pi \hbar\;.
\label{decay-equiv-width-ur}
\end{equation}
If the exponential decay \index{exponential decay}
formulas are inserted and the constant $x_{0}=t=0$
(for $f$), and $E=E_{0}$ (for $\tilde{f}$), then one obtains equality in the
above relation.

It is interesting to observe that the autocorrelation function describes
coherence in time. This is a useful measure of the \emph{fine structure }of
the temporal distribution function $\mathfrak{p}\left( t\right) =\left| f\left(
t\right) \right| ^{2}$.

A different approach to describing width and fine structure was taken by
Hilgevoord and Uffink (cf. the review of Hilgevoord \cite{Hil96,Hil98}), who
adopted the concepts of \emph{overall width} and \emph{translation width}
from the theory of signal analysis as follows. Let $\chi $ be a
square-integrable function, normalised to unity, and $\tilde{\chi}$ its
Fourier transform. The \emph{overall width }$\mathfrak{W}\left( \left| \chi
\right| ^{2},\alpha \right) $ of the distribution $\left| \chi \right| ^{2}$
is defined as the width of the smallest time interval $\mathfrak{\Theta }$ such
that \index{time--energy uncertainty relation}
\[
\int_{{\mathfrak\Theta}}\left| \chi \left( t\right) \right| ^{2}\,dt=\alpha\;.
\]
Then the following relation holds:
\begin{equation}
\mathfrak{W}\left( \left| \chi \right| ^{2},\alpha \right) \,\mathfrak{W}\left(
\left| \tilde{\chi}\right| ^{2},\alpha \right) \geq C\left( \alpha \right)
\,,\;\;\;\mathrm{for}\;\;\;\alpha >\frac{1}{2}\;,  \label{HU-ove-width-ur}
\end{equation}
with a constant $C\left( \alpha \right) $ independent of $\chi $. This
yields an energy time uncertainty relation in the spirit of the Wigner
relation (\ref{Wig-ur}) if we put $\chi \left( t\right) =f\left( t\right)
=\left\langle \varphi |\psi _{t}\right\rangle $, $\tilde{\chi}\left(
E\right) =\tilde{f}\left( E\right) $; in the case of $\varphi =\psi _{0}$
and $H$ having a nondegenerate spectrum, then $\tilde{f}\left( E\right)
=\left| \left\langle E|\psi _{0}\right\rangle \right| ^{2}$.

For the analysis of interference experiments, a relation between the overall
width of the energy distribution and the translation width of the temporal
distribution has proved enormously useful. The \emph{translation width }$%
\mathfrak{w}\left( f,\rho \right) $ is defined as the smallest number $t$ for
which
\[
\left| f\left( t\right) \right| =\left| \left\langle \psi _{0}|\psi
_{t}\right\rangle \right| =1-\rho \;.
\]
Then observing that $\tilde{f}\left( E\right) =\left| \left\langle E|\psi
_{0}\right\rangle \right| ^{2}$, Hilgevoord and Uffink \cite{HU88} show:
\index{Hilgevoord--Uffink relation}
\begin{equation}
\mathfrak{w}\left( f,\rho \right) \,\mathfrak{W}\left( \tilde{f},\alpha \right) \geq
2\hbar \arccos \left( \frac{2-\alpha -\rho }{\alpha }\right) \,,\;\;\;%
\mathrm{for}\;\;\rho \geq 2\left( 1-\alpha \right) \;.
\label{HU-trans-width-ur}
\end{equation}
The lifetime-linewidth relation \index{lifetime-linewidth relation}
is recovered for any decaying state by
putting $T_{1/2}=\mathfrak{w}\left( f,\sqrt{1/2}\right) $, $\alpha =0.9$, which
yields \cite{Hil98}:
\begin{equation}
T_{1/2}\,\mathfrak{W}\left( \tilde{f},0.9\right) \geq 0.9\hbar \;.
\label{HU-lifetime-ur}
\end{equation}
An interesting connection between the Mandelstam--Tamm relation
\index{Mandelstam--Tamm relation}
and the
Hilgevoord--Uffink relation
\index{Hilgevoord--Uffink relation} is pointed out in \cite{Uff}.

With this example we conclude our survey of intrinsic-time energy
relations, without any claim to completeness. For example a number of
rigorous results on the rate with which an evolving state `passes through' a
reference subspace are reported by Pfeifer and Frohlich \cite{Pfei,PF}. We
also recommend the recent reviews of Hilgevoord \cite{Hil96,Hil98} as a
lucid didactic account demonstrating the importance of the translation
width/overall width uncertainty relation in substantiating Bohr's rebuttal
of Einstein's attempts to achieve simultaneous sharp determinations of
complementary quantities.

\section{Quantum Clock\label{q-clock}}
\index{quantum clock}
\index{time--energy uncertainty relation}

The constituents of real rods and clocks and other measuring devices are
elementary particles, atoms and molecules, which are subject to the laws of
quantum mechanics. Hence it is natural to investigate the effect of the
quantum nature of measuring instruments. This thought has played a leading
role in the early debates between Einstein and the other founders of quantum
mechanics. By taking into account quantum features of the experimental
setup, Bohr was able to refute Einstein's \emph{Gedanken }experiments which
were aimed at beating quantum limitations of joint measurements of position
and momentum, or time and energy. Later Wigner exhibited limitations of
space-time measurements due to the quantum nature of test particles, and it
was in this context that he introduced the idea of a \emph{quantum clock }
\cite{Wig57,Sal}, see Chap. \ref{chap:clocks}.

The issue of quantum clocks \index{quantum clock}
belongs in a sense to the realm of the theory of
time measurements:\index{time measurement}
time is being measured by means of observing the
dynamical behaviour of a quantum system. However, the ensuing uncertainty
relations are clearly of the intrinsic-time type, and the theory of quantum
clocks is actually based on the theory of repeated measurements, or
monitoring, of a non-stationary quantum-nondemolition variable. By
contrast, time as an observable \index{time as an observable}
is recorded in experiments in which
typically a detector waits to be triggered by the occurrence of some \emph{%
event}, such as a particle hitting a scintillation screen. The latter type
of \emph{event time }measurement \index{event time}
will be discussed in the next section.

The Salecker--Wigner quantum clock \index{Salecker--Wigner clock}
has experienced renewed interest in
recent years in three areas of research: investigations on the detectability
of the quantum nature of spacetime on length scales far larger than the
Planck length (e.g. \cite{Ad,Ng}); studies of tunnelling times (e.g. \cite
{Leav}) and superluminal photon propagation through evanescent media \cite
{Japha}; and quantum information approaches to optimising quantum clock
resolution \cite{Buz} and synchronisation via nonlocal entangled systems
(e.g. \cite{Josza}). All of these questions and proposals are subject to
ongoing controversial scrutiny, so that it is too early to attempt an
assessment. Instead we will be content with a brief outline of the principal
features of a quantum clock and explain the relevance of the intrinsic--time
energy uncertainty relation in this context.

A quantum clock\index{quantum clock}
is characterised as a system that, in the course of its time
evolution, passes through a sequence of distinguishable states $%
\psi_{1},\psi_{2},\dots$ at (laboratory) times $t_{1},t_{2},\dots$ . In
order to be distinguishable as clock pointer positions, neighbouring states $%
\psi_{k},\psi_{k+1}$ must be (at least nearly) orthogonal. Under this
assumption, the time resolution defined by this system is $\delta
t=t_{k+1}-t_{k}$. It is known that a nonstationary state which runs through $%
n$ orthogonal states in a period $T$ must be a superposition of at least $n$
energy eigenstates. For a harmonic oscillator with frequency $\omega$ and
period $T=2\pi/\omega$, the state $\psi_{1}=\sum_{k=1}^{n}\varphi_{k}/\sqrt{n%
}$ will turn into $\psi_{2}$ perpendicular to $\psi_{1}$ if $\delta t=T/n$.
It follows that the mean energy must be of the order $\hbar/\delta t=\hbar
n/T$.\index{time--energy uncertainty relation}

If one considers the mean position of a wave packet as the clock pointer,
then according to the relevant Mandelstam--Tamm relation
\index{Mandelstam--Tamm relation}
and the constraint $%
\delta t>\tau _{\psi _{1}}\left( Q\right) $ on the resolution, one obtains
\[
\delta t\geq \frac{\hbar }{2\Delta _{\psi _{1}}H}\;.
\]
These examples illustrate the fact that the rate of change of a property of
the system decreases with increasing sharpness of the prepared energy. In
the limit of an energy eigenstate, all quantities will have time-independent
distributions and expectation values: hence \emph{nothing happens}.

Another requirement to be imposed on a system to ensure its functioning as a
quantum clock is that its pointer can be read in a non-disturbing way. This
can be achieved for suitable families of pointer states, such as coherent
states for the harmonic oscillator, which admit non-demolition measurements.
The relevant theory of quantum-nondemolition measurements for continuous
variables is developed in \cite{BL}.

The quantum clock time energy uncertainty relation can be derived in a very
general way from the intrinsic--time uncertainty relations reviewed above.
In order to achieve a time resolution $\delta t$, pairs of successive
pointer states $\psi_{1}=\psi_{t}$, $\psi_{2}=\psi_{t+\delta t}$ need to be
orthogonal: $\mathfrak{p}\left( \delta t\right) =\left| \left\langle \psi
_{t}|\psi_{t+\delta t}\right\rangle \right| ^{2}=0$. The relation (\ref{MT-p}%
) implies:
\begin{equation}
\delta t\geq t_{0}=\frac{\pi\hbar}{2\Delta_{\psi_{1}}H}\;.
\label{MT-clock-ur}
\end{equation}
As noted before, the variance is not always a good measure of the width of
the energy distribution. A more stringent condition on the clock resolution
can be obtained by application of the Hilgevoord--Uffink relation (\ref
{HU-trans-width-ur}) between temporal translation width and overall energy
width. If the clock is a periodic system, the resolution $\delta t$ is given
by the period divided by the number of pairwise orthogonal states, $\delta
t=T/n$. This entails that the state $\psi_{1}$ has to have a translation
width of the order of at most $\delta t$. Hence (\ref{HU-trans-width-ur})
yields:
\[
\delta t\geq\mathfrak{w}\left( f,\rho\right) \geq\frac{2\hbar\arccos\left(
\left( 2-\alpha-\rho\right) /\alpha\right) }{\mathfrak{W}\left( \tilde
{f},\alpha\right) }\;.
\]
For a quantum clock, $\rho$ should be close to unity. Taking $\rho=1$
requires $\alpha\geq\frac{1}{2}$, and we have
\[
\delta t\geq\frac{2\hbar\arccos\left( \left( 1-\alpha\right) /\alpha \right)
}{\mathfrak{W}\left( \tilde{f},\alpha\right) }\equiv\hbar\, C\left( \alpha\right)
\,,\;\;\;\frac{1}{2}\leq\alpha\leq1\;.
\]
Since both the enumerator as well as the denominator are increasing
functions of $\alpha$, and since the quotient $C\left( \alpha\right) $ is 0
both at $\alpha=\frac{1}{2}$ (as $\arccos1=0$) and at $\alpha=1$ (as $\mathfrak{W%
}\left( \tilde{f},1\right) =\infty$), it follows that there must be a value $%
\alpha_{0}$ where $C\left( \alpha\right) $ is maximal. The inequality for
the clock resolution must still hold at this point:
\begin{equation}
\delta t\geq\hbar C\left( \alpha_{0}\right) \;.  \label{HU-clock-ur}
\end{equation}
A universal quantum clock uncertainty relation in this spirit was proposed
by this author \cite{Bu90a} and independently by Hilgevoord and Uffink \cite
{HU91}.

\section{Relations Based on Time Observables\label{obs}}
\index{time--energy uncertainty relation}

Let us recall the motivation for considering time as a quantum observable.
\index{time as an observable}
First, there do exist a variety of experiments in which times of events are
recorded, where these events occur at randomly distributed instants as
monitored by means of laboratory clocks. The appropriate mathematical tool
for the representation of these temporal statistics is that of a POVM
\index{Positive Operator Valued Measure (POVM)}
over
the time domain, which will be explained in Subsection \ref{ev}. As an
illustration of intrinsic time preparation and measurement inaccuracies, we
will briefly review in Subsection \ref{box} the famous Einstein photon box
experiment. Secondly, having acknowledged the possible role of time as a
random variable, the next question that arises concerns the nature of the
randomness: for example, is the instant of decay of an unstable particle%
\emph{\ truly} indeterminate, as would be appropriate to a quantum
observable, or is it determined by some possibly hidden mechanism, albeit
unpredictable? We shall argue in Subsection \ref{t-interf} that an
indeterminacy interpretation is appropriate in the light of temporal
interference experiments.

\subsection{Event Time Observables\label{ev}}
\index{time--energy uncertainty relation}

A measurement of an ordinary quantum observable is typically devised so as
to provide an outcome at a specified instant of time. Often one aims at
achieving the \emph{impulsive measurement} limit where the duration of the
interaction between object and probe is negligible, so that it makes sense
to speak of an (approximate) instant of the measurement.

By contrast, event time \index{event time}
measurements are extended in time, with sensitive
detectors waiting to be triggered. The experimenter has no control over the
time instant at which the detectors fire. This very instant constitutes the
outcome of such a measurement.

Wigner \cite{Wig72} epitomises the distinction between these two types of
measurements in terms of the localisation of particles. The first type of
measurement amounts to measuring the position at a particular time. This
will answer the question: \emph{`Where is the particle -- now?'} The second
type of measurement corresponds to a determination of the instant of time at
which the particle passes a particular point in space, thus answering the
question: \emph{`When is the particle -- here?'}

Following \cite{Bu90a}, we explain the term \emph{event} to refer to the
(approximate) actuality of a property, in the sense that the probability for
this property to occur is equal to (or close to) unity. The event to be
observed in the above time of passage experiment is the approximate
localisation of the particle at the given space point. We note that the
Mandelstam--Tamm parameter $\tau _{\rho }\left( Q\right) $ seems to give an
indication of the \emph{indeterminacy} of the time of passage, owing to the
indeterminacy of position in the state $\rho$.\index{time of passage}

With the exception of the photodetection theory (e.g., \cite{Dav,Sri}), a
theory of \emph{event time} \index{event time}
measurements is very much in its initial stages.
In the 1990s, interest in the theory of \emph{time of arrival} measurements
has grown significantly and ensuing results are reviewed in other chapters
of this book. Here we focus on the formal representation of event time
observables in terms of POVMs.\index{time--energy uncertainty relation}
\index{Positive Operator Valued Measure (POVM)}

Suppose a detection experiment is repeated many times until a sufficiently
large statistical distribution of times is obtained. A quantum mechanical
account of the statistics will have to provide probabilities for the event
times to lie within intervals $Z$ of the time domain. Such probabilities
should be expressed as expectation values of operators associated with each
set $Z$, that is, $p_{\rho }\left( Z\right) =\mathrm{tr}\left[ \rho
\,F_{Z_{0}}\left( Z\right) \right] $. These probabilities should be
approximately equal to the observed frequencies. Here $Z_{0}$ denotes an
interval which represents the time domain specified in the experiment in
question. If the measurement can be thought of as being extended from the
infinite past to the infinite future, one would have $Z_{0}=\mathbb{R}$.

Due to the positivity of the numbers $p_{\rho}\left( Z\right) $ for all
states $\rho$, the operators $F_{Z_{0}}\left( Z\right) $ will be positive.
Similarly since $p_{\rho}\left( Z\right) \leq1$, we have $F_{Z_{0}}\left(
Z\right) \leq I$. Finally, the (countable) additivity of probability
measures entails the (countable) additivity of the $F_{Z_{0}}\left( Z\right)
$ for disjoint families of sets $Z_{k}$, that is, $F_{Z_{0}}\left( \cup
Z_{k}\right) =\sum_{k}F_{Z_{0}}\left( Z_{k}\right) $. Taken together, these
properties ensure that the family of $F_{Z_{0}}\left( Z\right) $ constitutes
a (not necessarily normalised) POVM over $Z_{0}$.
\index{Positive Operator Valued Measure (POVM)}
Due to the nature of time
measurements, one anticipates that certain events will never occur (i.e.,
for no state $\rho$), so that indeed it may happen that $p_{\rho}\left(
Z_{0}\right) <1$, or $F_{Z_{0}}\left( Z_{0}\right) <I$.

Every observable can be characterised by its transformation behaviour under
the fundamental space-time transformations. In particular, time observables
will transform covariantly \index{time covariance}\index{time as an observable}
under time translations:
\begin{equation}
U_{t}F_{Z_{0}}\left( Z\right) U_{t}^{-1}=F_{Z_{0}-t}\left( Z-t\right) \;.
\label{time-cov}
\end{equation}
Properties of such time observables and specific examples (mainly in the
context of decay observation) are studied in detail by Srinivas and
Vijayalakshmi \cite{Sri}. Detection times are axiomatically characterised as
\emph{screen observables} \index{screen observable}
through further transformation covariance
relations in work due to Werner \cite{Wer}.

Assuming that first and second moments for the POVM
\index{Positive Operator Valued Measure (POVM)}
$F_{Z_{0}}$ are defined
on a dense domain, one can introduce a unique maximally symmetric (generally
not self-adjoint) time operator \index{time--energy uncertainty relation}
\index{time operator}
\[
T=\int_{Z_{0}}t\,F_{Z_{0}}\left( dt\right) \;.
\]
We put $\overline{t}=\mathrm{tr}\left[ \rho\cdot T\right] $, then the
temporal variance is defined as
\begin{equation}\label{time-var}
\left( \Delta _{\rho }T\right) ^{2}=\frac{\int_{Z_{0}}\left( t-\overline{t}%
\right) ^{2}\,\mathrm{tr}\left[ \rho \,F_{Z_{0}}\left( dt\right) \right] }{%
\mathrm{tr}\left[ \rho \,F_{Z_{0}}\left( Z_{0}\right) \right] }\,\;.
\end{equation}
The uncertainty relation (\ref{1paul}) then follows for an event time
\index{event time} observable
and energy if the observation period $Z_{0}=\mathbb{R}$:
\begin{equation}\label{pov-ur}
\Delta _{\rho }T\,\Delta _{\rho }H\geq \frac{\hbar }{2}\;.  %\label{obs-ur}
\end{equation}
For an event time POVM
\index{Positive Operator Valued Measure (POVM)}
with a finite interval $Z_{0}$, this relation is not
generally valid.

It is still true, as it was in 1990 \cite{Bu90a}, that a systematic quantum
theory of time measurements \index{time measurement}
is lacking but will be necessary for an
operational understanding of event time POVMs. The following examples may serve
as guidance for the development of a better intuition about time
observables and measurements.
\index{time as an observable}

\subsubsection{Freely Falling Particle.}
\index{time--energy uncertainty relation}
\index{freely falling particle}

For the Hamiltonian
\begin{equation}
H_{g}=\frac{P^{2}}{2m}-mgQ\;,  \label{H-g}
\end{equation}
one easily verifies that the following self-adjoint operator $T_{g}$ is
canonically conjugate to $H$:
\begin{equation}
T_{g}=-\frac{1}{mg}P\;.  \label{T-g}
\end{equation}
In fact this choice is suggested by the dynamical behaviour of the system:
solving the Heisenberg equation of motion gives $P\left( t\right) =P-mgt\,I$,
where $P\left( 0\right) =P$. Time is measured dynamically as the linear
increase of momentum. In this case even the Weyl relations are satisfied:
\begin{equation}
e^{itH/\hbar }\,e^{ihT/\hbar }=e^{-ith/\hbar }\,e^{ihT/\hbar }\,e^{itH/\hbar
}\;.  \label{Weyl}
\end{equation}
As a further consequence, $T_{g}$ and $H$ act as generators of energy and
time shifts, respectively, in the sense of the covariance relations
\begin{eqnarray}
e^{ihT/\hbar }\,H\,e^{-ihT/\hbar }& =& H+hI\;,  \label{H-cov2} \\
e^{itH/\hbar }\,T\,e^{-itH/\hbar }& =& T-tI\;.  \label{T-cov}
\end{eqnarray}
The associated time POVM
\index{Positive Operator Valued Measure (POVM)}
is indeed a projection valued measure, namely, the
spectral measure
\[
E^{T_{g}}\left( Z\right) =E^{P}\left(- mgZ\right) \;.
\]
Both the covariance relations as well as the Weyl relation imply the
Heisenberg canonical commutation relation and hence the uncertainty
relation (\ref{pov-ur}).

It must be noted that the present Hamiltonian is unbounded, its spectrum
being absolutely continuous and extending over the whole real line. Thus the
obstruction due to Pauli's theorem \index{Pauli's ``theorem''}
does not apply.

\subsubsection{Oscillator Time.}
\index{time--energy uncertainty relation}

We now consider the Hamiltonian (putting $m=\hbar=1$)
\begin{equation}
H_{\mathrm{osc}}=\frac{1}{2}\left( P^{2}+Q^{2}\right) \;.  \label{H-osc}
\end{equation}
The spectrum consists of non-negative, equidistant values, so that there is
no unitary shift group, hence no self-adjoint operator $T$ satisfying the
Weyl relation (\ref{Weyl}) can exist. Nevertheless, classical reasoning
suggests the existence of a phase-like quantity that transforms covariantly
(modulo 2$\pi$) under the time evolution group. This leads to the
introduction of a time POVM
\index{Positive Operator Valued Measure (POVM)}
and hence a periodic time variable proportional
to the phase.

Introduce the ladder operator $a=\frac{1}{2}\left( Q+iP\right) $, which
gives the number operator $N=a^{\ast}a$, with eigenvalues $n=0,1,2,\dots$
and eigenvectors $|n\rangle$. Then $H=N+\frac{1}{2}I$. For $t\in\left[ 0,2\pi%
\right] $, we introduce the formal, non-normalisable vectors $%
| t\rangle=\sum_{n}e^{int}\,|n\rangle$, then we define:
\[
F_{\mathrm{osc}}\left( Z\right) =\left( 2\pi\right) ^{-1}\,\int
_{Z}dt\,|t\rangle\langle t|=\sum_{n,m\geq0}\left( 2\pi\right) ^{-1}\int
_{Z}e^{i\left( n-m\right) t}\,dt\,|n\rangle\langle m|\;.
\]
It is easily verified that this defines a normalised, shift covariant (mod 2$%
\pi$) POVM.
\index{Positive Operator Valued Measure (POVM)}

This oscillator-time POVM yields a whole family of self-adjoint operators
canonically conjugate to $H_{\mathrm{osc}}$: first define
\[
T_{\mathrm{osc}}^{\left( 0\right) }=\int_{0}^{2\pi}t\,F_{\mathrm{osc}}\left(
dt\right) =\sum_{m\neq n\geq0}\frac{1}{i\left( n-m\right) }%
\,|n\rangle\langle m|\,+\pi I\;.
\]
This operator was first constructed as a self-adjoint solution of
the canonical commutation relation (\ref{ccr}), thus refuting a
widespread erroneous reading of Pauli's theorem.\index{Pauli's
``theorem''} Consequently, this operator does satisfy the
uncertainty relation (\ref{pov-ur}) in a dense domain (certainly not
containing the energy eigenstates). Strangely enough, this aspect of
the interesting papers of Garrison and Wong \cite{GarW} and Galindo
\cite{Gal85} has been widely ignored, while the fact as such is
repeatedly being rediscovered in recent years. Next we calculate the
time shifts of this operator,
\[
T_{\mathrm{osc}}^{\left( t\right) }=e^{itH}\,T_{\mathrm{osc}}^{\left(
0\right) }\,e^{-itH}=T_{\mathrm{osc}}^{\left( 0\right) }-tI+2\pi F_{\mathrm{%
osc}}\left( \left[ 0,t\right] \right)\;.
\]
Here we are facing a covariant family of non-commuting, self-adjoint
operators, all of which satisfy the canonical commutation relation with $%
H=H_{\mathrm{osc}}$. The non-commutativity corresponds to the fact that the
phase quasi-eigenvectors $|t\rangle$ are mutually non-orthogonal, so that $%
F_{\mathrm{osc}}$ itself turns out to be a non-commutative POVM.
\index{time--energy uncertainty relation}
\index{Positive Operator Valued Measure (POVM)}

We have here given just one example of a covariant oscillator time (phase)
POVM. There are in fact an infinite variety of such phase POVMs associated
with $H_{\mathrm{osc}}$. First significant steps towards a systematic
account and operational analysis of covariant oscillator phase POVMs have
been recently undertaken by Lahti and Pellonp\"{a}\"{a} \cite{LahP99}.
\index{Positive Operator Valued Measure (POVM)}

We note that a similar construction to the present one is possible for a
finite quantum system with a spin Hamiltonian
\[
H_{\mathrm{spin}}=\beta s_{3}\;,
\]
where $s_{3}$ is the $z$ component of the spin of a spin--$s$ system.
However, in this case a canonical commutation relation and Heisenberg
uncertainty relation are not valid.

\subsubsection{Time POVMs vs. Time Operators?}

The preceding example shows in a striking way that observables may
be more appropriately represented by means of a POVM instead of just
a self-adjoint or symmetric operator: not only does the latter
merely give the first moments of the experimental statistics, but,
as seen here, there may exist a high degree of non-uniqueness in the
choice of even a self-adjoint operator as a representative of an
observable (here the phase, or oscillator time). An approach to
defining event time observables taking into account the
characteristic covariance may help to remove these ambiguities.

Nevertheless, for specific systems for which the physics of time
measurements is well understood, the construction of  canonical time
operators may be sufficient and adequate.

By providing some mathematical qualifications
on\index{Pauli's theorem} Pauli's claims concerning self-adjoint
time operators canonically conjugate to the Hamiltonian of a
physical system, Galapon \cite{Gal2002a,Gal2002b} made room for the
construction of such \index{time operator} canonical time operators
for certain positive Hamiltonians with non-empty point spectrum.
This was recently followed with a fresh approach to the time of
arrival operator for a free particle in
\cite{Gal2004,Gal2005a,Gal2005b}; see also Chapter 10. In the next
example we provide some general considerations on  the search for
covariant POVMs corresponding to the time of arrival.

\subsubsection{Free Particle Time Observables.}
\index{time--energy uncertainty relation}

Seemingly obvious candidates of a time operator conjugate to
the free particle Hamiltonian,
\[
H_{\mathrm{free}}=\frac{P^{2}}{2m}\;,
\]
are given by suitably symmetrised expressions for the time-of-arrival variable
suggested by classical reasoning, such as, for example:
\[
-\frac{1}{2}m\left( QP^{-1}+P^{-1}Q\right)
\qquad {\rm or}\qquad -m P^{-1/2}QP^{-1/2}\, .
\]
\index{time operator} While these expressions formally satisfy the canonical
commutation relation, they are \emph{not self-adjoint } but only symmetric (on
suitably defined dense domains on which they actually coincide, see Sec.~10.4), and
they do not possess a self-adjoint extensions. Hence this intuitive approach
does not lead to a time observable \index{time as an observable} in the usual
sense of a self-adjoint operator conjugate to the free Hamiltonian. For a long
time, this observation has been interpreted by many researchers as implying
that time is not an observable in quantum mechanics. But this view does not
take into account the fact that there are detection experiments which record
the time of arrival of a particle, or more precisely, the time when the
detector fires. The statistics of such measurements are appropriately described
as probability distributions using suitable POVMs. For the present case of a
free particle there do indeed exist time-shift covariant, normalised POVMs. An
example is given by the following:
%\begin{equation*}
%\langle\varphi|F_{\mathrm{free}}\left( Z\right) \varphi\rangle=\left(
%2\pi\right) ^{-1}\int_{Z}dt\,\left| \int_{\mathbb{R}}dp\sqrt{\left| p\right|
%/m\hbar}\,\exp\left( itp^{2}/m\hbar\right) \,\tilde{\varphi }\left( p\right)
%\right| ^{2}\;.
%\end{equation*}
\begin{eqnarray*}
\langle\varphi|F_{\mathrm{free}}\left(  Z\right)  \varphi\rangle&=&\left(
2\pi\right)  ^{-1}  \int_{Z}dt\,\bigg\{\left|
\int_{0}^{\infty}dp\sqrt{p/m\hbar
}\,\exp\left(  itp^{2}/m\hbar\right)  \,\tilde{\varphi}\left(  p\right)
\right|  ^{2}+\\
&&\qquad\qquad+\left|  \int_{-\infty}^{0}dp\sqrt{-p/m\hbar}\,\exp\left(  itp^{2}%
/m\hbar\right)  \,\tilde{\varphi}\left(  p\right)  \right|  ^{2}\bigg\}\;.
\end{eqnarray*}
%The construction of such time POVMs is based on the Neumark extension of
%symmetric operators such as $T_{\mathrm{free}}$ to a self-adjoint operator
%in a larger Hilbert space, for which a spectral measure does exist; the POVM
%is then obtained as the projection of the projection valued measure to the
%original Hilbert space.
Early explicit constructions of such POVM time observables
\index{time as an observable}
and more general
\emph{screen} observables can be found in \cite{Hol} and \cite{Wer}.
More recently, the question of constructing time of flight observables
as covariant POVMs has been intensely studied; this development is
reviewed in Chap. \ref{chap:toa}.
\index{Positive Operator Valued Measure (POVM)}
\index{time--energy uncertainty relation}

\subsubsection{Time POVM associated with an effect.}

The question of defining a \index{time operator}time observable for
any given type of event was investigated by Brunetti and Fredenhagen
\cite{BruFre2002a} who were able to define a time translation
covariant \index{Positive Operator Valued Measure (POVM)}POVM associated
with a positive operator representing the event in question
(an \emph{effect}, in the terminology of Ludwig \cite{Lud1983}).
These authors also derived a new lower bound for the time
uncertainty for covariant event time POVMs on the time domain $\mathbb
R$, \cite{BruFre2002b}:
\begin{equation}
\Delta_\rho T\ge \frac{d}{\langle H\rangle_\rho}
\end{equation}
Using their approach, Brunetti and Fredenhagen were able to rederive
the time delay operator of scattering theory. This work has
inspired new model investigations on the theory of time measurements
\cite{Hegetal2002,Hegetal2003}.

In order to illustrate Brunetti and Fredenhagen's approach, we construct
a simple example of a covariant time POVM associated with a
Hamiltonian $H$ with simple bounded, absolutely continuous spectrum
$[0,2\pi]$. One can think of a particle moving in one spatial dimension, with its
momentum confined to the interval $[0,p_0]$, where $p_0^2/2m=2\pi$.

Let $\mathcal{H}$ be the Hilbert space $L^2(0,2\pi)$ in which
$H$ acts as the multiplication operator $H\psi(h)=h\psi(h)$. We
choose a shift-covariant family of unit vectors $\varphi_t$, $t\in\mathbb R$,
as follows (putting
$\hbar=1$): $\varphi_t(h)=e^{iht}/\sqrt{2\pi}$. We can then define a time-shift
covariant POVM via
\begin{equation}\label{time-povm}
P(X):=\int_X|\varphi_t\rangle\langle\varphi_t|\,dt,\quad
X\in\mathcal{B}(\mathbb{R}).
\end{equation}
The normalization $P(\mathbb{R})=I$ can be verified by considering
the integral
\[
\int_{\mathbb{R}}\langle\psi|\varphi_t\rangle\langle\varphi_t|\xi\rangle\,dt.
\]
for any $\psi,\xi\in\mathcal{H}$, and showing that its value is
$\langle\psi|\xi\rangle$. This follows readily by observing that the
function
$
t\mapsto\langle\varphi_t|\xi\rangle=:\hat\xi(t)
$
is the
Fourier-Plancherel transform $\hat\xi=:\mathcal{F}\xi$ of $\xi\in
\mathcal{H}$. Note that $\hat\xi\in L^2(\mathbb{R})$, and that
$\mathcal{F}(\mathcal{H})$ is a proper closed subspace  of
$L^2(\mathbb{R})$. Thus we find that for $\psi\in\mathcal{H}$,
\[
\mathcal{F}P(X)\mathcal{F}^{-1}\hat\psi(t)=\chi_X(t)\hat\psi(t),
\]
which corresponds to the Naimark extension of the POVM $P$ to a spectral
measure on $L^2(\mathbb R)$.

We are now in a position to compare the time observable (\ref{time-povm}) with
the general construction of  Brunetti and Fredenhagen in
\cite{BruFre2002a}. Given a bounded positive operator $A$, they consider
the positive operator measure, defined first on intervals $J$ via
\[
B(J):=\int_J e^{itH}Ae^{-itH}\,dt.
\]
They then show that in certain circumstances this can be turned into a normalized POVM
on a suitable closed subspace (provided this is not the null space). In the present case
of the POVM (\ref{time-povm}), we see that the operator corresponding to $A$ can be identified
with the 1-dimensional projection operator $|\varphi_0\rangle\langle\varphi_0|$. In that case the normalization condition is already satisfied, and $B(J)=P(J)$
holds on $\mathcal{H}$. The POVM $P$ corresponds to a measurement of the time that the system
spends (loosely speaking) in the state $\phi_0$.

A formal time operator is obtained from the first moment operator of the
POVM $P$:
\begin{equation}
T\psi(h)=\int_{\mathbb{R}}t\,\varphi_t(h)\langle\varphi_t|\psi\rangle\,dt
=-i\frac{d}{d h}\psi(h);
\end{equation}
this is well-defined for functions $\psi\in L^2(0,2\pi)$ which are absolutely continuous
and such that the derivative $\psi'\in L^2(0,2\pi)$. In order for this operator to be symmetric,
the domain must be further restricted by appropriate boundary conditions. It is well known
that the condition $\psi(2\pi)=c\psi(0)$ makes $-id/dh$ a self-adjoint operator
$T^{(c)}$ for any $c$ of modulus 1. Each such $T^{(c)}$ is a self-adjoint extension of
the differential operator understood as a symmetric operator $T^{(0)}$ with the boundary condition
$\psi(0)=\psi(2\pi)=0$. Note that the spectrum of $T^{(c)}$ is $\mathbb Z$, with eigenvectors
$e^{i\arg(c)H/2\pi}\varphi_m$, where $\varphi_m(h)=e^{imh}/\sqrt{2\pi}$, $m\in\mathbb Z$.

The covariance relation
\[
e^{i\tau H}Te^{-i\tau H}=T-\tau I
\]
is found to be satisfied for $T^{(0)}$ but not for any of its self-adjoint extensions since
$e^{i\tau H}T^{(c)}e^{-i\tau H})=T^{(c')}$ with $c'=e^{i2\pi\tau}c$.  In accordance with this,
the canonical commutation relation between the Hamiltonian and the time operator is obtained only on the domain of $T^{(0)}$,
and therefore the uncertainty relation (\ref{pov-ur}) holds on this dense
subspace, with the  variance of the time distribution being defined
via equation (\ref{time-var}). Since the spectrum of $H$ is a bounded interval of length
$\lambda(H)=2\pi$, there is an absolute bound to the temporal variance in any state $\rho$:
\begin{equation}
\Delta_\rho T \ge \frac\hbar{2\lambda(H)}.
\end{equation}

These examples show that for a variety of Hamiltonians, event time
\index{event time}
observables can be defined as time-shift covariant POVMs, the form of which
is inferred by the aid of classical intuition or with reference to a
class of experimental situations. Where the first moment
operator satisfies a canonical commutation relation with the Hamiltonian on
a dense domain, the observable-time energy uncertainty relation will follow.
Whether or not this is the case depends on the nature of the spectrum of the
Hamiltonian and the time domain \cite{Sri}.

We conclude this brief survey of the problem of time-covariant POVMs
with the following pointers to some interesting related
developments.

A connection between time observables represented by POVMs and
irreversible dynamics has been explored by Amann and Atmanspacher
\cite{AA1997}.

Finally, there have been several studies of the representation of
event time observables in terms of POVMs in the wider context of
relativistic quantum mechanics and quantum gravity
\cite{Gia1,Gia2,Tol1,Tol2,Maz,Rov}. It is too early and beyond the
scope of the present chapter to give a conclusive review of these
recent and ongoing developments.

\subsection{Einstein's Photon Box\label{box}}
\index{Einstein's photon box}
\index{time--energy uncertainty relation}

A comprehensive theory of event time \index{event time}
measurements is missing to date, so
that a first step towards an understanding of time as an observable seems to
be to carry out case studies. Here we will revisit briefly the \emph{%
Gedanken }experiment proposed by Einstein. In this experiment, a photon is
allowed to escape from a box through a hole which is closed and opened
temporarily by a shutter. The opening time period is determined by a clock
which is part of the box system. Einstein argued that it should be possible
to determine the energy of the outgoing photon by weighing the box before
and after the opening period. Thus it would seem that one can obtain an
arbitrarily sharp value for the energy of the photon, while at the same
time the time period of preparation, or emission of the outgoing photon
could be made as short as one would wish, by setting the clock mechanism
appropriately. This conclusion would contradict the preparation-time energy
uncertainty relation (\ref{prep-ur}).

Bohr's rebuttal \cite{Bohr48} was based on the observation that the accuracy
of the weighing process is limited by the indeterminacy of the box momentum,
which in turn limits the unsharpness of position by virtue of the Heisenberg
uncertainty relation for the box position and momentum. But an uncertainty
in the box position entails an uncertainty in the rate of the clock, as a
consequence of the equivalence principle. All this taken together, the
accuracy of the determination of the photon energy and the uncertainty of
the opening time do satisfy the uncertainty relation (\ref{1paul}).

Bohr's informal way of reasoning has given rise to a host of attempts, by
some, to make the argument more precise (or even more comprehensible) or, by
others, to refute it in defence of Einstein. In fact if Bohr's were the only
way of arguing, the consistency of nonrelativistic quantum mechanics
(replacing the photon with a (gas) particle) would appear to depend on the
theory of relativity. Hence several authors have considered different
methods of measuring the photon energy.
\index{time--energy uncertainty relation}

In his review of 1990, the present author has offered an argument that makes
no assumptions concerning the method of measurement and is simply based on a
version of quantum clock uncertainty relation. This argument goes as
follows. If the photon energy is to be determined with an inaccuracy $\delta
E$ \ from the difference of box energies before and after the opening
period, then these energies must be well defined within $\delta E$, that is,
the box energy uncertainty $\Delta E$ must satisfy $\Delta E\leq\delta E$.
Then the clock uncertainty relation, either in the Mandelstam--Tamm form (\ref
{MT-clock-ur})
\index{Mandelstam--Tamm relation}
or the Hilgevoord--Uffink form (\ref{HU-clock-ur}),
\index{Hilgevoord--Uffink relation}
allows us
to conclude that the box system needs at least a time $t_{0}\cong\hbar/%
\Delta E$ in order to evolve from the initial `shutter closed' state to the
(orthogonal!) `shutter open' state (and back). During this transition time $%
t_{0}$ it is \emph{objectively indeterminate }whether the shutter is open or
closed. Accordingly, also the time interval within which the photon can pass
the shutter is indeterminate by an amount $\Delta T=t_{0}$. We thus arrive
precisely at Bohr's relation
\begin{equation}
\Delta T\,\delta E\cong\hbar\;.  \label{box-ur}
\end{equation}
It seems satisfying that this derivation works without advocating the box
position-momentum uncertainty relation; instead it refers directly to the
quantum dynamical features of the box. Without going into an analysis of the
energy transfer between box and photon, it seems plausible that the energy
measurement uncertainty $\delta E$ of the box, which corresponds to an
uncertainty of the box energy, will give rise to an uncertainty of the
energy of the escaping photon. Similarly, the uncertainty in the shutter
opening time \ gives a measure of the uncertainty of the time of passage of
the photon through the hole. Hence the box uncertainty relation admits also
the following interpretation: it is impossible to determine the energy and
time of passage \index{time of passage}
of a particle with accuracies better than those allowed by
this uncertainty relation. Thus the measurement uncertainty relation (\ref
{box-ur}) accords with the dynamical Mandelstam--Tamm relation
\index{Mandelstam--Tamm relation}
for the
characteristic time $\tau_{\rho}\left( Q\right) $, equation (\ref{MT-ur}),
and thus with the preparation-time relation (\ref{prep-ur}).
\index{time--energy uncertainty relation}

It is also interesting to note the close analogy between this experiment and
the double slit experiment where similar debates between Bohr and Einstein
took place concerning the possibility of jointly determining the position
and momentum of a particle. Time of passage and energy are complementary
quantities in the same sense as position and momentum: the arrangements for
determining time (position) and energy (momentum) are mutually exclusive.
However, while these conclusions have been corroborated in the case of
position and momentum with appropriate quantum mechanical joint measurement
models (for details and a survey of this development, cf. \cite{BGL95}), a
similarly comprehensive treatment for time and energy is as yet waiting to
be carried out. Only very recently a first scheme of joint measurements of
energy and time of arrival \index{time of arrival}
has been proposed \cite{Muga2} along the lines of
the position-momentum measurement model due to Arthurs and Kelly.
\index{Arthurs--Kelly model}

\subsection{Temporal Interference and Time Indeterminacy\label{t-interf}}
\index{temporal interference}
\index{time indeterminacy}
\index{time--energy uncertainty relation}

In the preceding sections we have repeatedly referred to temporal
indeterminacies of events such as the passage of a particle through a space
region, and we have motivated this interpretation indirectly by invoking the
quantum indeterminacies of the relevant dynamical properties. The analogy
between the time--energy complementarity and the position-momentum
complementarity that emerges in the context of the\index{Einstein's
photon box} Einstein photon box (a
point strongly emphasised by Cook \cite{Cook}) suggests, however, that it
should be possible to obtain direct experimental evidence for the
appropriateness of the indeterminacy interpretation of time uncertainties.
In the case of position and momentum, the indeterminacy of the location of a
particle passing through a screen with two slits is demonstrated by means of
the interference pattern on the capture screen which images the fine
structure of the associated momentum amplitude function. As a simple model
illustration, if the wave function of the particle at the location of the
slit is given as
\[
\psi_{0}\left( x\right) =\left\{
\begin{array}{lll}
\left( 4a\right) ^{-1/2} &  & \mathrm{if\;\;\;}A-a\leq\left| x\right| \leq
A+a\;, \\
&  &  \\
0 &  & \mathrm{elsewhere\;,}
\end{array}
\right.
\]
then the momentum amplitude is given as the Fourier transform,
\[
\tilde{\psi}_{0}\left( p\right) =2\sqrt{a}\,\cos\left( Ap\right) \,\frac{%
\sin\left( ap\right) }{ap}\;.
\]
If the slit width $a$ is small compared to the distance between the slits
then the factor $\left( \sin\left( ap\right) /ap\right) ^{2}$ describes the
slowly varying envelope of the momentum distribution while the factor $%
\cos^{2}\left( Ap\right) $ describes the rapid oscillations that constitute
the interference pattern. If the path of the particle were known, one would
have an incoherent mixture of two packets travelling through the slits, and
no interference would appear. Hence the ignorance interpretation regarding
the two paths is in conflict with the presence of the interference pattern
which is due to the coherent superposition of the two path states. In other
words, the path is indeterminate, and it is \emph{objectively undecided}
through which slit the particle has passed.
\index{time--energy uncertainty relation}

In a similar way, if one were able to offer a particle a multiple \emph{%
temporal }`slit', then the indeterminacy of the time of passage
\index{time of passage}would be
reflected in an interference pattern in the associated energy distribution.
As it turns out, experiments exhibiting such \emph{diffraction in time},
\index{diffraction in time} or
\emph{temporal interference}, \index{temporal interference}
had already been carried out in the 1970s. In
the experiment of Hauser, Neuwirth and Thesen \cite{HNT}, a beam of M\"{o}%
ssbauer quanta is emitted from excited $^{57}$Fe nuclei, with a\ mean energy
of $E_{0}=$14.4 keV and a lifetime $\tau =141$ ns, and is sent through a
slit which is periodically closed and opened by means of a fast rotating
chopper wheel. Then the energy distribution of the quanta is measured. The
count rate is around 3000 events per second, so that on average there is
about one photon within 2000 lifetimes passing the device. This suggests
that one is observing interference of individual photons. We briefly sketch
the analysis and interpretation proposed in \cite{Bu90a}.

The amplitude incident at the chopper,
\[
f_{0}\left( t\right) =e^{-t/2\tau}\,e^{-i\omega_{0}t}\,,\;\;\;\omega
_{0}=E_{0}/\hbar,\,\ \ \ \ \ t\geq0\;,
\]
is modulated into
\[
f\left( t;t_{0}\right) =f_{0}\left( t\right) \,\chi\left( t;t_{0}\right) \;.
\]
Here the chopping function $\chi$ is equal to one for all $t>0$ which fall
into one of a family of equidistant intervals $Z_{k}$ of equal length $T_{%
\mathrm{open}}$ distributed periodically, with period $T_{\mathrm{chop}}$,
over the whole real line. For all other values of $t$ we have $\chi\left(
t\right) =0$. The time parameter $t_{0}$ indicates the difference between
the zero point of the decay process and the beginning of a chopping period;
its value is distributed uniformly over a chopping period if a large
ensemble of events is observed.

\index{time--energy uncertainty relation}
The Fourier transform of $f_{0}$ reproduces the Lorentzian shape of (\ref
{f-til-E-Lor}). The energy distribution obtained behind the chopper should
be given by the Fourier transform of $f\left( t;t_{0}\right) $,
\[
\tilde{f}\left( \omega ;t_{0}\right) =\int_{\mathbb{R}}dt\,f\left(
t;t_{0}\right) e^{i\omega t}=\sum_{k}\int_{Z_{k}}dt\,\,f_{0}\left( t\right)
\,e^{i\omega t}=\sum_{k}\tilde{f}_{k}\left( \omega ;t_{0}\right) \;.
\]
Hence, the expected spectral intensity is
\begin{equation}
\mathfrak{I}\left( \omega ;t_{0}\right) =\left| \tilde{f}\left( \omega
;t_{0}\right) \right| ^{2}=\left| \sum_{k}\tilde{f}_{k}\left( \omega
;t_{0}\right) \right| ^{2}\;.
\end{equation}
This corresponds to a coherent superposition of the temporal partial packets
$f_{k}\left( t;t_{0}\right) $. The observed distribution is obtained by
averaging $\mathfrak{I}\left( \omega ;t_{0}\right) $ over one chopping period
with respect to $t_{0}$,
\begin{equation}
\mathfrak{I}\left( \omega \right) =\frac{1}{T_{\mathrm{chop}}}\int_{0}^{T_{%
\mathrm{chop}}}\,dt_{0}\,\mathfrak{I}\left( \omega ;t_{0}\right) \;.
\end{equation}
Now, if one assumed the time window through which each photon passes to be
\emph{objectively }determined (albeit possibly unknown), then one would
predict the $t_{0}$-average $\mathfrak{I}^{\mathrm{ob}}\left( \omega \right) $
of the spectral distribution
\begin{equation}
\mathfrak{I}^{\mathrm{ob}}\left( \omega ;t_{0}\right) =\sum_{k}\left| \tilde{f}%
_{k}\left( \omega ;t_{0}\right) \right| ^{2}\;.
\end{equation}
A calculation yields that the shape of the distribution $\mathfrak{I}^{\mathrm{ob%
}}\left( \omega \right) $ is very similar to a somewhat broadened Lorentzian
curve, whereas $\mathfrak{I}\left( \omega \right) $ shows a sharp central peak
and several distinguished, symmetric side peaks of much smaller amplitudes.
The latter is in excellent agreement with the experimental spectral data.
\index{time--energy uncertainty relation}

The increase of the overall width of the spectral distribution can be seen
as a consequence of the temporal fine structure introduced by the action of
the chopper. Similarly, the fine structure of the spectral distribution is
linked to the overall width of the temporal distribution: the latter is of
the order of the lifetime, \index{lifetime}
while the former is approximately equal to the
undisturbed linewidth. This behaviour is in accordance with the
Hilgevoord--Uffink relation \index{Hilgevoord--Uffink relation}
(\ref{HU-trans-width-ur}) between overall width
and translation width for a pair of Fourier-related distributions, which is
thus found to be (at least qualitatively) confirmed.

We conclude that the spectral interference pattern exhibited in this
experiment demonstrates the \emph{non-objectivity}, or \emph{indeterminacy}\
of the time of passage \index{time of passage}
of the photon through the chopper. It is tempting to
go one step further and claim that the time of the emission of the photon is
equally indeterminate.

In 1986, time indeterminacies were demonstrated for material particles, in
an observation of \emph{quantum beats} \index{quantum beats}
in neutron interferometry by Badurek
et al \cite{Badur}. Similar temporal diffraction experiments
\index{temporal diffraction experiments}have been
carried out in recent years with material particles, namely atoms \cite{Sz}
and neutrons \cite{Hils}. The results obtained are in agreement with the
time energy uncertainty relation. The issue of the (non-)objectivity of
event times \index{event time}
has also been investigated from the perspective of Bell's
inequalities. In a seminal paper of Franson \cite{Fran}, an interference
experiment with time--energy entangled photons
\index{time--energy entangled photons} was proposed. Subsequent
measurements by Brendel et al \cite{Bre93} and Kwiat et al \cite{Kwiat}
yielded observed fringe visibilities in accordance with quantum mechanical
predictions and significantly larger than allowed by a Bell inequality
\index{Bell inequality} that
follows from classical reasoning.
\index{time--energy uncertainty relation}

\section{Conclusion}

We summarise the main types of time energy uncertainty relations
\begin{equation}
\Delta T\,\Delta E\gtrsim \hbar
\end{equation}
and their range of validity depending on the interpretation of the
quantities $\Delta T$ and $\Delta E$:

(1) A relation involving\emph{\ external time} is valid if $\Delta T$ is the
\emph{duration }of a perturbation or preparation process and $\Delta E$ is
the uncertainty of the energy in the system.

(2) There is \emph{no }limitation to the duration of an energy measurement
and the disturbance or inaccuracy of the measured energy.

(3) There is a variety of measures of \emph{characteristic, intrinsic times}%
, with ensuing \emph{universally }valid \emph{dynamical time energy
uncertainty relations}, $\Delta E$ being a measure of the width of the
energy distribution or its fine structure. This comprises the Bohr-Wigner,
Mandelstam--Tamm, Bauer--Mello, and Hilgevoord--Uffink relations.
\index{Mandelstam--Tamm relation}
\index{Hilgevoord--Uffink relation}

(4) \emph{Event time observables} can be formally represented in terms of
positive operator valued measures over the relevant time domain. An \emph{%
observable-time energy uncertainty relation}, with a constant positive lower
bound for the product of inaccuracies, is \emph{not universally} valid but
will hold in specific cases, depending on the structure of the Hamiltonian
and the time domain.

(5) Time measurements by means of \emph{quantum clocks} are subject to a
dynamical time energy uncertainty relation, where the time resolution of the
clock is bounded by the unsharpness of its energy, $\delta t\gtrsim
\hbar/\Delta E$.
\index{time--energy uncertainty relation}

(6) Einstein's photon box experiment constitutes a demonstration of the
\emph{complementarity} of time of passage and energy: as a consequence of
the quantum clock uncertainty relation, the inaccuracy $\delta E$ in the
determination of the energy of the escaping photon limits the uncertainty $%
\Delta T$ of the opening time of the shutter. This is in accordance with the
\emph{energy measurement }uncertainty\emph{\ }relation based on \emph{%
internal clocks }discovered recently by Aharonov and Reznik.

(7) Temporal diffraction experiments provide evidence for the \emph{%
objective indeterminacy }of event time uncertainties such as time of passage.
\index{time of passage}

Finally we have to recall that:

(8) A full-fledged quantum mechanical theory of time measurements is
still waiting to be developed.

\vspace{12pt}

\noindent {\bf Acknowledgement.} Work leading to this revised and expanded version was carried out
during the author's stay at the Perimeter Institute, Waterloo, Canada. Hospitality and support
by PI is gratefully acknowledged.

%\clearpage
%\addcontentsline{toc}{section}{Index}
%\flushbottom
%\printindex
%%%%%%%%%%%%%%%%%%%%%%%%%%%%%%%%%%%%%%%%%%%%%%%%%%%%%%%%%%%%%%%%%%%%%
\end{document}